\documentclass[useAMS,usenatbib,usegraphicx]{mn2e}

\title[Carbon monoxide in the atmosphere of Pluto]
{Discovery of carbon monoxide in the upper atmosphere of Pluto}
\author[J. S. Greaves et al.]{J. S.
Greaves$^{1}$\thanks{E-mail: jsg5 at st-andrews.ac.uk}, Ch. 
Helling$^1$ \& P. Friberg$^{2}$\\
$^{1}$SUPA, Physics \& Astronomy, University of St Andrews, North
Haugh, St Andrews, Fife KY16 9SS, UK\\
$^{2}$Joint Astronomy Centre, 660 North A`oh\={o}k\={u} Place, 
University Park, Hilo, HI 96720, USA
}

\begin{document}

\date{Accepted 2011. Received 2011; in original form 2011}

\pagerange{\pageref{firstpage}--\pageref{lastpage}} \pubyear{2011}

\maketitle

\label{firstpage}

\begin{abstract}

Pluto's icy surface has changed colour and its atmosphere has 
swelled since its last closest approach to the Sun in 1989. The 
thin atmosphere is produced by evaporating ices, and so can also 
change rapidly, and in particular carbon monoxide should be 
present as an active thermostat.  Here we report the discovery of 
gaseous CO via the 1.3mm wavelength J=2-1 rotational transition, 
and find that the line-centre signal is more than twice as bright 
as a tentative result obtained by Bock\'{e}lee-Morvan et al. in 
2000. Greater surface-ice evaporation over the last decade could 
explain this, or increased pressure could have caused the 
atmosphere to expand. The gas must be cold, with a narrow 
line-width consistent with temperatures around 50~K, as predicted 
for the very high atmosphere, and the line brightness implies that 
CO molecules extend up to $\approx 3$ Pluto radii above the 
surface. The upper atmosphere must have changed 
markedly over only a decade since the prior search, and more 
alterations could occur by the arrival of the New Horizons mission 
in 2015.

\end{abstract}


\section{Introduction}

Pluto was the first body to be discovered in the Kuiper Belt of 
icy planetesimals orbiting in the outer Solar System. Although now 
only the second-largest body known in the belt (after Eris), Pluto 
is intensively studied, and is the primary target of the New 
Horizons mission \citep{stern03}, with a spacecraft flyby 
scheduled in 2015. In particular, Pluto offers the rare 
opportunity to study a very cold planetary atmosphere which also 
evolves strongly with time, and may in fact be transient, settling 
and freezing out on the surface \citep{owen93} as Pluto recedes 
from the Sun in its highly elliptical orbit. As Pluto's last 
perihelion was in 1989, subsequent contraction of the atmosphere 
was expected, but in fact when observed by occultation of stars 
the atmospheric pressure and size had increased between 1988 and 
2002 \citep{elliot03a}. Around the same time, the surface ice had 
become redder, and also brighter around the northern pole as it 
came more into sunlight \citep{buie10a,buie10b}. Many of these 
results were surprising, such as changes in albedo over as little 
as a year, and darkening of the southern regions, contrary to 
expectations if ices have been sublimating from here most 
recently.

The atmosphere and surface must be strongly coupled, since 
sublimation of ices supplies the atmosphere 
\citep[e.g.]{schaller07}, and so as the composition of the surface 
changes seasonally \citep{buie10a,buie10b}, the gas abundances 
should also evolve. Atmospheric temperatures of around 100~K 
\citep{elliot03b} are much warmer than the surface at only 40 K 
\citep{tryka94}. This requires a complex thermal balance 
controlled by the relative abundances of molecules that absorb 
radiation and re-emit in cooling lines, plus there is an overall 
influence of the level of sunlight, particularly in the 
ultra-violet \citep{strobel07}. The atmosphere is also very 
fragile, with a solar-driven slow wind predicted to be removing it 
into space \citep[e.g.]{delamere09}.

\begin{figure*}
\label{fig1}
\includegraphics[width=180mm,angle=0]{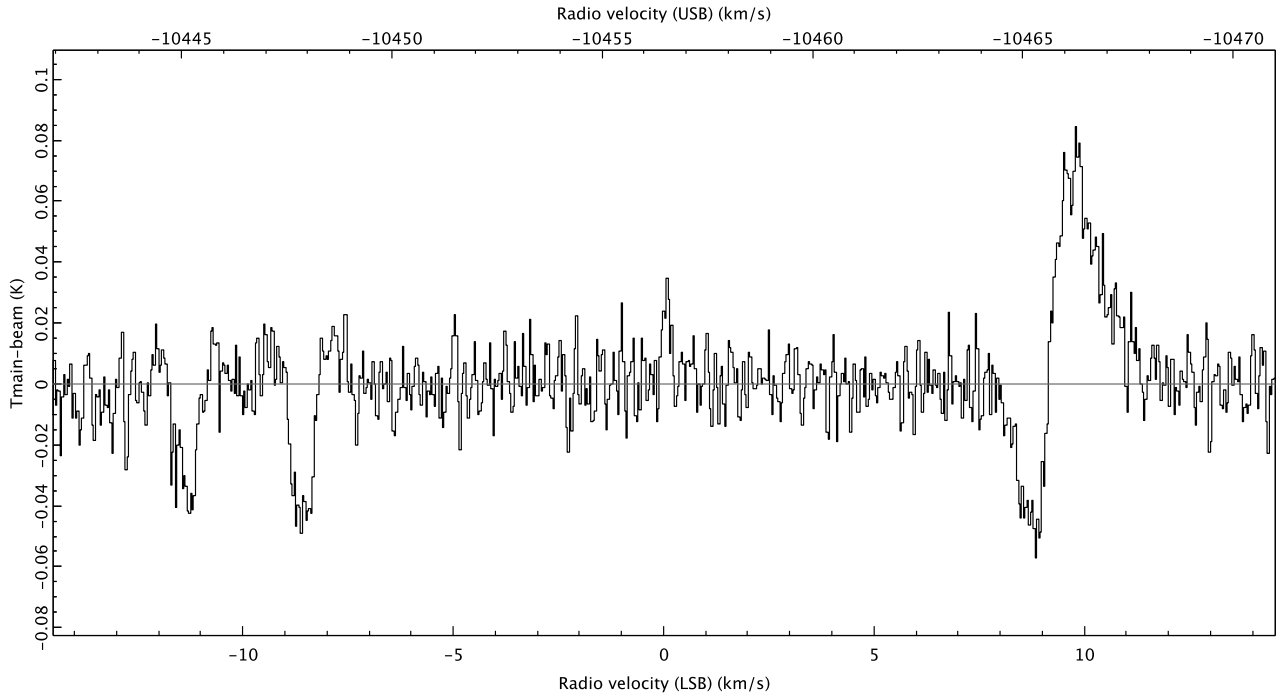}
\caption{Spectrum towards Pluto in the J=2-1 rotational
transition of CO at 1.3mm wavelength. The atmospheric line 
from Pluto appears near zero velocity (in the dwarf planet's
rest-frame) while other lines are residuals from background
Galactic clouds varying from night to night. Summed over eight
spectral channels, the Pluto line is detected at 6.5$\sigma$.}
\end{figure*}

Such a cold diffuse atmosphere is difficult to study by 
remote observation, and most conclusions are presently 
model-based. The trace gas methane is the only one to have 
been confirmed spectroscopically \citep{young97}. Surface-ice 
spectroscopy shows frozen nitrogen dominates \citep{owen93}, 
with methane and carbon monoxide also present, plus possibly 
ethane \citep{demeo10} and nitriles \citep{protopapa08}. 
N$_2$ must dominate the atmosphere as it vaporizes readily 
and is by a factor of 50 the most abundant ice 
\citep{owen93}, but sublimated CO is the strongest coolant 
\citep{strobel07} and so should act as the thermostat. Some 
evidence for CO includes the brightest patch on Pluto's 
surface, suggested to be CO ice \citep[and references 
therein]{buie10b}, and an extinction layer seen in 
atmospheric occultation, interpreted as droplets of N$_2$ or 
CO \citep{rannou09}.

\subsection{Previous searches for CO} 

Prior searches for gaseous CO have been made with ground-based 
telescopes. Infrared absorption spectroscopy has recently 
constrained CO:N$_2$ to be $<$0.5~\% near the surface 
\citep{lellouch10}, an order of magnitude below an earlier limit 
\citep{young01}. There is also a long history of 
millimetre-wavelength searches, looking for rotational transitions 
of CO from the higher atmosphere \citep{barnes93,bockelee01}. As 
the atmosphere is warmer than the surface, and extends beyond the 
planetary disc, emission lines are predicted. The major difficulty 
is that Pluto is small and distant, subtending less than a percent 
of the beam solid angle, even for telescopes of 10-30m class. An 
upper limit was first obtained using the Haystack 37m in 1993, for 
the CO J=1-0 transition rotational line at 2.6 mm wavelength 
\citep{barnes93}. This limit was improved by the equivalent of 
eight-fold in 2000, with the IRAM 30m telescope 
\citep{bockelee01}, and this group also searched for the J=2-1 
transition at 1.3~mm with better sensitivity. A tentative positive 
signal was seen, with the brightest spectral channel observed at 
28 mK in main-beam brightness temperature, versus 1$\sigma$ noise 
of 10 mK per 0.1~km/s spectral channel. However, the observation 
was severely affected by a strong line from a background Galactic 
cloud, coming as close as 0.5 km/s to potential Pluto emission 
\citep{bockelee01}.

The 2000 data suffered from an unfortunate coincidence in sky 
position with an uncatalogued Galactic source, but Pluto's orbit 
as seen from the Earth has brought it closer to the Galactic Plane 
over the last couple of years. Here we present results from the 
15m James Clerk Maxwell Telescope, making a new search for the 
J=2-1 line, at epochs chosen to minimise Galactic contamination 
behind Pluto. The search was motivated by time availability at an 
excellent millimetre-site, and the opportunity to examine the 
evolution of Pluto's atmosphere in the exciting run-up period to 
the arrival of New Horizons \citep{stern03}. We report the 
confirmation of atmospheric CO (after two decades of searches), 
making it only the second gas-phase species to be detected, after 
methane \citep{young97}.

\section{Observations and reduction}

\begin{figure*}
\label{fig2}
\includegraphics[width=125mm,angle=0]{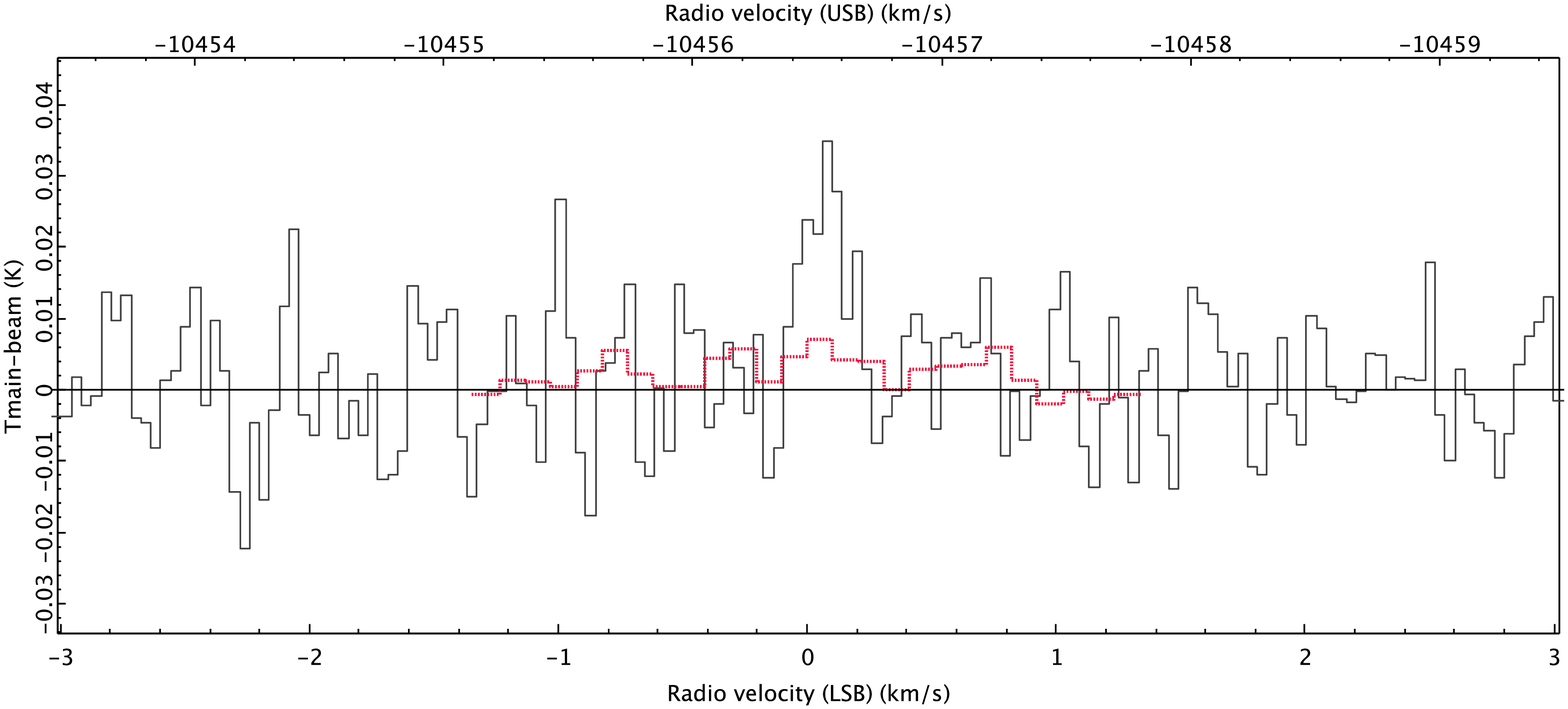}
\includegraphics[width=48mm,angle=0]{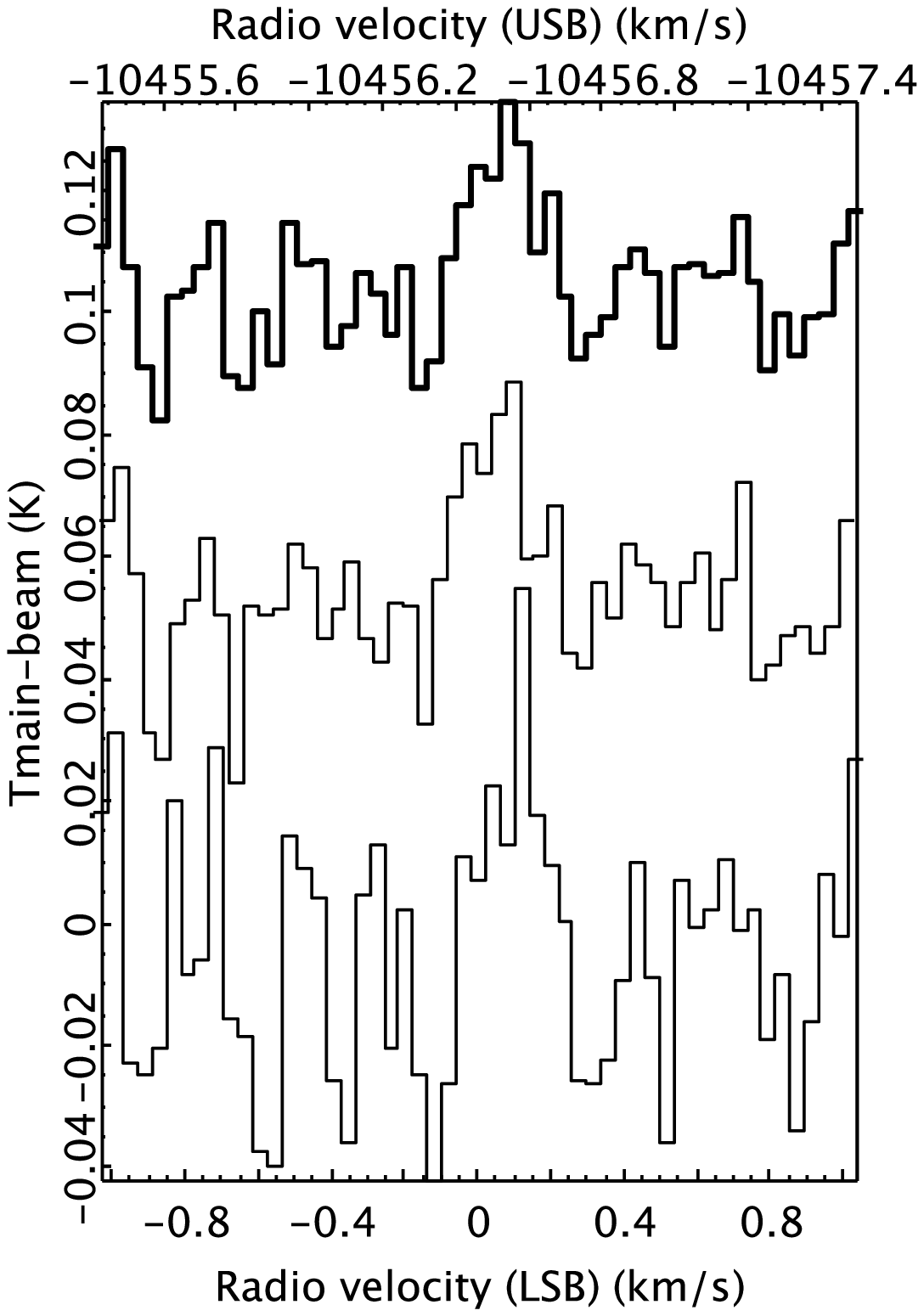}
\caption{Left: expanded version of Figure~1. Overplotted in red 
is the spectrum obtained in April 2000 by \citet{bockelee01} 
using the IRAM 30m, but re-scaled for equivalent signal with 
the 15m JCMT; the brightest channel near zero velocity is then 
$7\pm2.5$~mK over 0.1 km/s. Right: comparison of the average 
spectrum and the 2010 and 2009 data (top to bottom, offset for 
clarity). 
}
\end{figure*}

\begin{figure}
\label{fig3}
\includegraphics[width=85mm,height=60mm,angle=0]{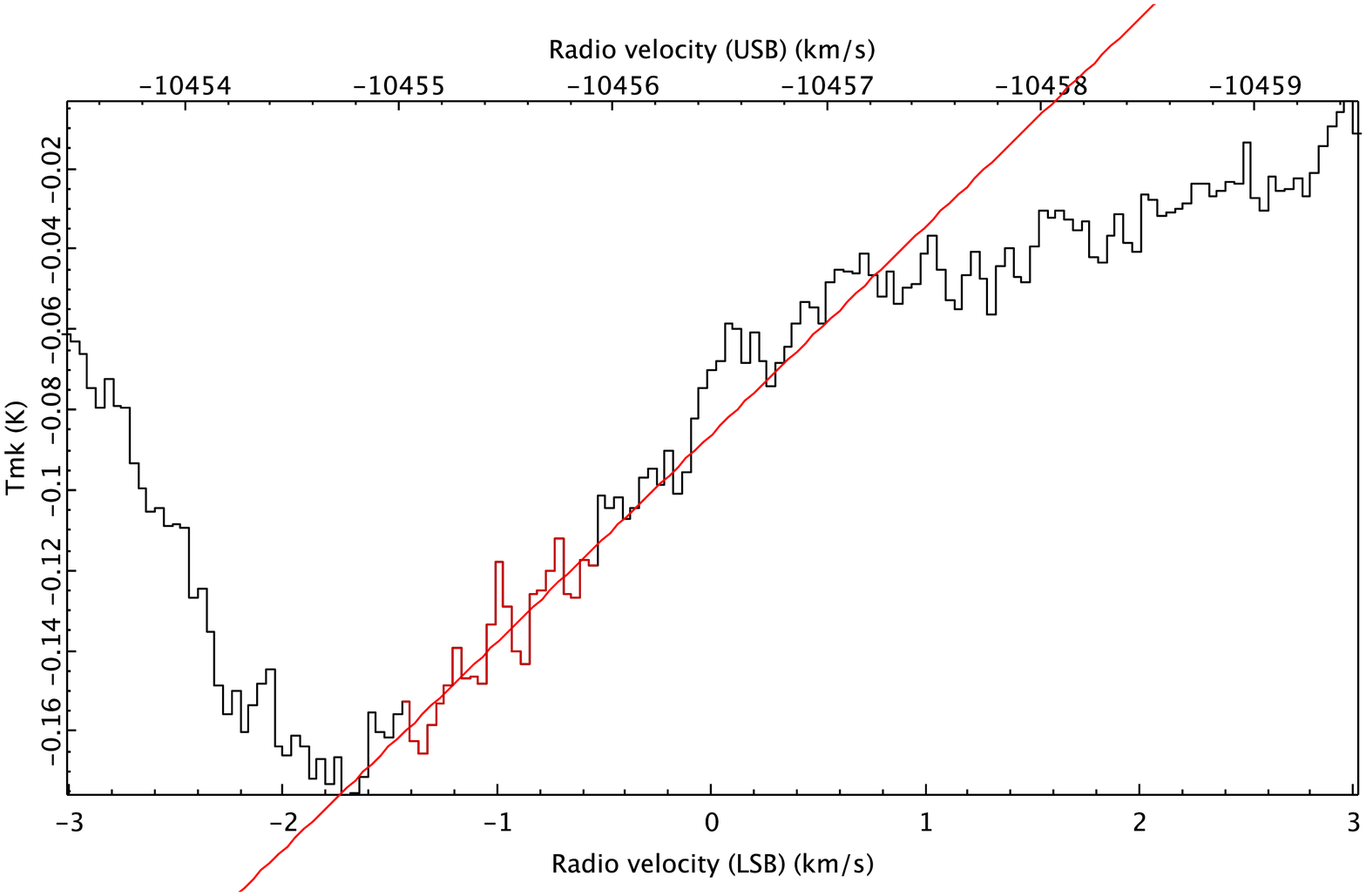}
\caption{Example spectrum if all 11 nights of data are weighted 
and co-added but no baselines are removed. The red channels were 
used to fit an example linear baseline, and this would extrapolate 
across the Pluto velocities as shown by the red straight line.
}
\end{figure}

We used the JCMT at 4000 m altitude on Mauna Kea, Hawaii, 
observing over 3 nights in August 2009 and 8 nights in April to 
May 2010. Conditions were good to average for the site, with 
zenith sky opacity at 1.3~mm typically around 0.1. The setup used 
the ACSIS spectrometer \citep{buckle09} together with the A3 
receiver \citep{cunningham92} (the instrument described was 
subsequently rebuilt by the Herzberg Institute of Astrophysics 
with a 1.3~mm-band mixer). Total on-source integration times were 
12 and 19 hours in 2009 and 2010 respectively, with individual 
observations lasting approximately 20 minutes. The observing mode 
was frequency-switching, with a throw of 35.5 km/s within the 
passband. The telescope tracked Pluto's sky position but not its 
velocity, so spectra were shifted in software to the dwarf 
planet's reference frame. The spectrometer channel spacing was set 
to 0.04~km/s and there are systematics of around $\pm$0.025~km/s 
for velocity-drifts of Pluto within each observation that have not 
been corrected. Data are shown on the main-beam antenna 
temperature brightness scale, with a beam efficiency of 0.75 at 
230.538 GHz frequency.

For each night of data, the shifted spectra were co-added, and 
then the appropriate positive and negative frequency-switched 
features were averaged. The region of the net positive line was 
blanked and replaced with a linear interpolation of surrounding 
channels. A running average over 12 channels was then generated 
from this and subtracted from the pre-blanked spectrum as a 
baseline. The 11 nightly spectra made in this way were finally 
combined, weighted by 1/noise$^2$ factors to take into account 
their different conditions and observing durations. Rejected data 
include two earlier nights using a smaller frequency-switch, and 
two later nights when Pluto had approached within 1.5$^{\circ}$ of 
the Galactic Plane. These spectra show ripples and complex 
baselines, degrading the final line profile. The weight of these 
rejected data represent only 15~\% of the total observations made.


The observations were planned to maximise the height of Pluto from 
the Galactic Plane, which varies as seen from the Earth, to 
minimise contamination problems. Pluto's Galactic latitude in 2009 
was +2.2$^{\circ}$, and in 2010 was -2.0$^{\circ}$ to 
-1.7$^{\circ}$, with longitudes of $\approx +12^{\circ}$. Using 
two periods well spaced around the year (Aug and Apr/May) also 
caused large changes in Pluto's velocity relative to the 
telescope, from +26 km/s in 2009 to -27 to -19 km/s in 2010. Thus 
any narrow line seen persistently at Pluto's velocity in the 
co-added data can only be from the planetary atmosphere, as any 
Galactic background emission is different every night. Galactic 
lines were seen around Pluto's velocity in the spring data, but 
were much broader than the Pluto line.

\section{Results}

Figure~1 shows the co-added Pluto spectrum from 2009/10, in the 
dwarf planet's rest-frame. A clear line is seen at zero velocity, 
with 6.5$\sigma$ confidence in the integrated antenna temperature. 
There are no other significant narrow features; the next brightest 
(at -5 km/s), is at only 2.7$\sigma$ over three spectral channels, 
and was traced to a signal from a pair only of the 11 observing 
nights. The broad features are residuals from the running-average 
subtraction process, arising from broad lines of Galactic clouds 
from different nights. A detail spectrum (Figure~2) shows the very 
flat baseline achieved near Pluto's velocity, while a test made by 
blanking a broader region (20 channels instead of 12) showed a 
change in integrated brightness of only around 10~\%. This argues 
for a robust signal measurement of the narrow line seen, and also 
implies that broad line wings are faint.

The CO line from Pluto was independently detected in both the
2009 and 2010 epochs, at levels of 3.9$\sigma$ and 5.4$\sigma$
respectively, and with the same integrated brightness within the
errors (Figure~2, right panel). The antenna temperatures are 
calibrated to about 20~\%, based on rms scatter of the line-peak 
signal of spectral standards taken every night on the nearby 
source OH17.7-2. Also, inclusion of the two nights of early data 
taken with a smaller frequency-switch was found to reduce the 
final integrated line intensity by $\approx 20$~\%. The brightness
measurements could be affected by broad structure in the
baseline, and Figure~3 demonstrates the spectrum obtained if the
data are co-added without baselining. The curvature due to
Galactic residuals makes it difficult to establish the nature of 
faint line wings.

The integrated antenna temperature of CO J=2-1 over 8 spectral 
channels (0.32 km/s) is 6.3$\pm$1.0 mK km/s on a main-beam 
brightness temperature scale. The mean $T_{mb}$ is 20~mK, and the 
1$\sigma$ noise per 0.04~km/s spectral channel is 8.6 mK (twice as 
deep as \citet{bockelee01} for a common spectral resolution). The 
full-width at half-maximum intensity is approximately 0.28~km/s. 
The velocity centroid is $\approx +0.07 \pm 0.055$~km/s, 
including errors from comparing adjacent half-line 
intensities of 0.03~km/s and systematics of the 0.04~km/s channel 
spacing and $\pm0.025$~km/s velocity drifts within scans. Hence, 
the centroid velocity may be marginally positively offset from 
Pluto.

Figure~2 also shows the CO J=2-1 spectrum obtained in 2000 by 
\citet{bockelee01}. Their $T_{mb}$ values have been scaled 
downwards by a factor of four, to account for the greater 
beam-dilution for a compact source with the JCMT 15m versus the 
IRAM 30m telescope. On this `JCMT-equivalent' scale, the 2000 
spectrum has a 3$\sigma$ upper limit of 7.5 mK in each 0.10 km/s 
channel, and the brightest channel near zero velocity is 7~mK. 
Weak positive signals were seen in line wings, and the integrated 
signal over $\pm1.3$~km/s was 5.9 mK km/s, at signal-to-noise 
ratio of 4.5. The integrated intensity appears similar in 2000 and 
2009/10, but the earlier spectrum has a greater contribution from 
possible line wings (Figure~2). A Voigt-profile fit to the new 
spectrum suggests broad wings now contribute $\la35$~\%.

\subsection{Features of the CO spectrum}

Figure~2 shows that the line-centre emission in 2009/2010
significantly exceeds that in the observation from April 2000
\citep{bockelee01}. In the brightest 0.1 km/s channel, the
signal was $7\pm2.5$~mK, compared to $27\pm5$~mK now when
averaged over the same region from the JCMT spectrum. Thus the
ratio is at least 22/10.5 from the $1\sigma$ bounds, implying
the line-centre intensity has more than doubled over a decade.

The core of the line now seen is narrower than previously 
predicted in models \citep{stansberry94,strobel96}. Model line 
profiles as plotted by \citet{bockelee01} are over twice as wide 
as observed, with full-width half-maxima of approximately 
0.7~km/s. These models had atmospheric temperatures of 80-106 K, 
but our measured line width of 0.28 km/s implies that thermalised 
CO can not be warmer than $\approx 50$~K. This estimate includes 
only thermal broadening, and so the gas could be even cooler if 
there is another source of velocity dispersion. \citet{person08} 
find evidence for large-scale atmospheric waves, but with 
horizontal wind speeds of under 3~m/s, while upwards sublimation 
flows should be at only a few cm/s \citep{toigo10}. The latter 
would also be in the contrary sense to the tentative red-shift 
noted above, since upwelling would produce a blue-shift from the 
near-side of the atmosphere that is preferentially observed.

\section{Discussion}

The marked increase in CO line-peak brightness could be due to 
recent sublimation of surface ice -- for example, the largest 
bright patch seen on Pluto's surface is attributed to carbon 
monoxide frost \citep{buie10b}. While this is a long-term feature 
in albedo maps, it could be a rich source of sublimating CO. 
Alternatively, a large expansion in isobar heights was detected 
between 1988 and 2002 \citep{elliot03a}, and such increases in 
atmospheric pressure could raise the effective emitting area for 
trace molecules. However, these occultation data only detect the 
atmosphere up to around 135~km \citep[e.g.]{young08}, so effects 
at higher altitudes are uncertain. Regarding timescales, marked 
surface changes have occurred over intervals as short as a year 
\citep{buie10b}, while vertical sublimation speeds \citep{toigo10} 
imply a molecule could rise thousands of kilometres within a 
decade. Hence it is reasonable that temporal changes in gas 
densities and/or abundances as surface ices sublimate could result 
in the CO lines being quite different over observations a decade 
apart (Figure~2).

Finding cold CO points to gas at very large heights -- for example 
temperatures as low as 50~K have been predicted for about 4500 km 
above the centre of Pluto \citep{strobel07}. This is under solar 
minimum conditions, as actually present in 2009/10. If the top of 
the CO layer is at 4500~km, then the atmosphere would subtend a 
diameter of 0.4 arcsec, as it was seen in 2009/10 with Pluto at 
31~AU from the Earth. From this, in the simplest approximation of 
a filled opaque shell of CO molecules thermalised at 50 K, the 
beam-filling factor would be $3.6 \times 10^{-4}$, giving an 
expected line brightness of 18~mK. This is in good agreement with 
our mean $T_{mb}$ of 20~mK. Lower atmospheric layers should be 
warmer and so add line brightness, but as an example, gas twice as 
hot (100~K) is predicted to lie below 2000~km \citep{strobel07}, 
and thus be five times more beam-diluted.

\subsection{Comparison to models}

Exact calculations for the CO line profile require radiative 
transfer calculations and an atmospheric model, beyond the scope 
of this Letter. However, some estimates can be made for the 
heights of the emitting layers. The density profile should decline 
approximately as a power-law \citep{strobel07}, if CO has a 
constant mixing ratio with N$_2$. Then the height below which the 
CO emission becomes opaque can be estimated, or more strictly, the 
height of the tangent where the column of CO molecules excited to 
the J=2 level produces $\int \tau dv \approx 1 \times \Delta v$, 
for opacity $\tau$ of unity and $\Delta v$ equal to the line FWHM 
of 0.28 km/s. In an isothermal 50~K approximation, this height is 
2750~km for a power-law density profile approximating that at 
solar maximum \citep{strobel07}, and 2250~km for a solar minimum 
profile. The respective heights which contribute opacity of 0.01 
are about 4600 and 3200~km. These values are for CO:N$_2$ of 
0.005, following \citet{strobel07} and at an upper limit observed 
in July 2009 \citep{lellouch10}. If contributions to the CO signal 
come from as high up as 4500~km, as estimated above from 
beam-filling arguments, then it appears that atmospheric densities 
must be at the high end of the models, or upper-level CO must be 
more abundant than assumed. Also, in the models high densities are 
associated with warmer temperatures than infered here -- this 
anomaly may be related to heating from CH$_4$ molecules  
absorbing solar UV, noting that recent estimates of the methane 
mixing ratio \citep{zalucha11} are up to $\sim5$ times lower than 
assumed by \citet{strobel07}.


The CO upper boundary estimated to be at about 4500~km above the 
centre of Pluto corresponds to $\approx 4$ radii, for $R_{\rm 
Pluto} \approx 1153$~km \citep{tholen97}. This is a reasonable 
upper bound, given that above $\approx$~3-5 R$_{\rm Pluto}$, the 
gas should become so diffuse that is is collisionless 
\citep{strobel07}, and thus the CO J=2 level will not be 
populated. Also, interaction with the solar wind will cause 
ionised gas (and any swept-up neutrals) to flow away into space, 
above $\approx5$ R$_{\rm Pluto}$ in the simulations of 
\citet{delamere09,harnett05}. The CO layer estimated to extend out 
to $\approx 4$ R$_{\rm Pluto}$ is thus inside this flow zone and 
can be expected to be stable. \citet{strobel07} models a slower 
hydrodynamic outflow inside this region, but predicted speeds 
only exceed our 0.04~km/s velocity resolution above 4500~km, so 
effects on the spectrum would be negligible. It is however 
possible that the marginal CO line red-shift, if real, could 
indicate a flow forming into a comet-like tail directed away from 
the Sun.

CO gas should be chemically stable even at large altitudes, and at 
low densities, cloud models imply that it will stay in the gas 
phase. CO ice could only form if seed particles existed -- akin to 
noctilucent clouds on Earth, where highly supersaturated water can 
condense out on rare solid particles, such as meteoritic dust 
raining in from space. Such interplanetary dust is however sparse 
out in the Kuiper Belt \citep{landgraf02}. \citet{rannou09} 
proposed cloud decks lower down in Pluto's atmosphere, with 
droplets of CO or N$_2$ that could explain an extinction layer 
\citep{elliot03a}. However, heterogeneous nucleation (i.e. 
chemical reactions on a surface leading to increase of the droplet 
volume) is only efficient inside a limited temperature interval, 
and it yields no droplets above heights of $\sim$30 km in these 
Pluto models \citep{rannou09}.

Since surface changes \citep{buie10a,buie10b} and the gaseous 
methane abundance \citep{zalucha11} are seen to be temporally 
variable, time-dependent models will be needed to balance the 
effects of heating, cooling, sublimation and solar irradiation. 
Further CO line monitoring is underway, and N$_2$ gas can be 
detected for the first time with ultraviolet spectroscopy 
\citep{stark07} by the ALICE spectrograph on New Horizons 
\citep{stern08}. Hence atmospheric models can be tested from data 
taken before and during the 2015 flyby.

\section{Conclusions}

Carbon monoxide is confirmed in the atmosphere of Pluto, and gives 
the first detection of high-altitude gas. There have been large 
temporal changes in the CO line profile over the last decade, in 
parallel to major alterations of the surface ices. Pluto's rapidly 
changing and fragile atmosphere could provide an interesting 
test-bed for models of global climates.

\section*{Acknowledgments} 


This work was supported by the Scottish Universities Physics
Alliance. The data were obtained under programme M09BU12 at the
JCMT, with invaluable assistance from I. Coulson and J. Hoge.
The James Clerk Maxwell Telescope is operated by the Joint
Astronomy Centre on behalf of the Science and Technology
Facilities Council of the United Kingdom, the Netherlands
Organisation for Scientific Research, and the National Research
Council of Canada.

\bsp

\label{lastpage}

\end{document}